\documentclass[published]{JHEP3} 
\JHEP{00(2009)000}

\JHEPspecialurl{http://jhep.sissa.it/JOURNAL/JHEP3.tar.gz}

\usepackage{epsfig,multicol,bbm}

\title{
Constraints on neutrino masses from WMAP5 and BBN in the lepton
asymmetric universe 
}

\author{Maresuke Shiraishi \\
  Department of Physics and Astrophysics, Nagoya University, Aichi 464-8602, Japan \\ 
Email: mare@a.phys.nagoya-u.ac.jp}
\author{Kazuhide Ichikawa \\ 
Department of Micro Engineering, Kyoto University, Kyoto 606-8501, Japan \\ 
Email: kazuhide@me.kyoto-u.ac.jp}
\author{Kiyotomo Ichiki \\ 
Department of Physics and Astrophysics, Nagoya University,
Aichi 464-8602, Japan \\ 
Email: ichiki@a.phys.nagoya-u.ac.jp}
\author{Naoshi Sugiyama \\ 
Department of Physics and Astrophysics, Nagoya University, Aichi 464-8602, Japan \\ 
Institute for the Physics and Mathematics of the Universe (IPMU), The University of Tokyo, Kashiwa, Chiba, 277-8568, Japan \\ 
Email: naoshi@a.phys.nagoya-u.ac.jp}
\author{Masahide Yamaguchi 
\\ Department of Physics and Mathematics, Aoyama Gakuin
University, Sagamihara 229-8558, Japan \\
Email: gucci@phys.aoyama.ac.jp}

\received{\today}
\accepted{\today}

\preprint{0904.4396}


\abstract{
 In this paper, we put constraints on neutrino properties such as mass
 $m_{\nu}$ and degeneracy parameters $\xi_i$ from WMAP5 data and light
 element abundances by using a Markov chain Monte Carlo (MCMC) approach. In
 order to take consistently into account the effects of the degeneracy
 parameters, we run the Big Bang Nucleosynthesis code for each value of
 $\xi_i$ and the other cosmological parameters to estimate the Helium
 abundance, which is then used to calculate CMB anisotropy spectra
 instead of treating it as a free parameter. We find that the constraint
 on $m_{\nu}$ is fairly robust and does not vary very much even if the
 lepton asymmetry is allowed, and is given by $\sum m_\nu < 1.3 \ \rm
 eV$ ($95 \% \rm C.L.$).
}

\begin{document}

\section{Introduction}

Neutrino masses are the key feature beyond the standard model of
particle physics because neutrinos are assumed to be massless in the
minimal standard model. However, oscillation experiments recently
suggest tiny but non-zero masses of neutrinos. Unfortunately they can
probe the mass (squared) differences between different mass eigenstates
but cannot determine absolute neutrino masses.
In addition, the mass of electron-type neutrino is bounded as $m_{\nu_e} < 2.0$~eV (95\% C.L.) from the current laboratory experiment of tritium beta-decay \cite{tritium}. However, the most stringent
constraint on the sum of neutrino masses comes from cosmological
observations. For example, 
by using the SDSS luminous red galaxies \cite{SDSSred}, the sum of
neutrino masses is constrained as $\sum 
m_{\nu} < 0.9$~eV (95\% C.L.), though there is still uncertainty in the
galaxy bias. On the other hand, Ichikawa et al.\cite{mnu,mnu3} showed
that similar constraint can be obtained by CMB data alone, which is free
from such an uncertainty and the most conservative. The latest WMAP
results yield $\sum m_{\nu} < 1.3$~eV (95\% C.L.) for the $\Lambda$CDM
model \cite{komatsu}. This constraint is shown to be robust over
the different cosmological models, even if we
abandon the assumption of the spatial flatness of the universe, change
the equation of state of dark energy, and include tensor modes
\cite{mnu,mnu3,komatsu,dunkley}.

Another assumption in the standard cosmology that is usually taken is
the symmetry in the lepton sector. This is closely related to
cosmological neutrinos since neutrinos have all the lepton number of the
universe due to its electronic charge neutrality. The standard
sphaleron process forces the amount of lepton asymmetry to be the same
order as that of baryon asymmetry.  However, several ways are proposed
to avoid such sphaleron effects
\cite{EKM,FTV,Shi,CCG,Mcdonald,qball,yamaguchi,CTY,TY,LS}. For example,
if $Q$-balls are formed after the Affleck-Dine leptogenesis, the large
lepton asymmetry inside $Q$-balls is protected from sphaleron effects
and the unprotected small lepton asymmetry is converted into the small
baryon asymmetry \cite{qball}. Thus, small baryon asymmetry is naturally
compatible with large lepton asymmetry. In addition, observational
constraint on lepton asymmetry is much weaker than that of baryon
asymmetry. Then, one may wonder how robust the constraint on neutrino
masses is in the possible presence of large lepton asymmetry. This is
the main topic we would like to pursue in this paper.
   
The lepton asymmetry which is usually denoted by the degenerate
parameter $\xi$ is an interesting quantity to be constrained
\cite{Dolgovreview,SR,SS,popa}, where $\xi$ is defined as a chemical
potential between neutrino and anti-neutrino normalized by the neutrino
temperature. Recently, Popa and Vasile have obtained 
a limit on $\xi$ by using the WMAP 5 year data combined with the other
CMB data and the LSS data \cite{popa}. However, since there are uncertainties
of biasing and non-linearity as regards galaxy clustering data, it would
be useful to constrain $\xi$ from WMAP5 alone.  
We also consider indirect effects of $\xi$ 
on CMB through the
primordial helium abundance $Y_p$ generated during the process of the
Big Bang Nucleosynthesis (BBN).

In this analysis, we consider three light neutrinos, based on the result
of high energy experiments
such as LEP, i.e. $N_\nu = 2.984 \pm 0.008$
\cite{lep}. We assume equal neutrino masses between flavors, such as
$m_\nu \equiv m_{\nu_e} = m_{\nu_\mu} = m_{\nu_\tau}$, which yields
$\sum m_\nu = 3 m_\nu$. It is because the accuracy of the present
cosmological observation is not yet up to specify their mass differences. 
Potentially, the degeneracy parameter $\xi$ can take different values for
each flavor of neutrinos. In this paper we consider the two cases:
one is the case that 
$\xi_e=\xi_\mu=\xi_\tau$, and the other is that
$\xi_\mu=\xi_\tau\equiv \xi_{\mu,\tau}\neq\xi_e$. 
The first case
is motivated by Refs. \cite{dolgov,wong,ABB}, which show that if a flavor
mixing angle of neutrino $\theta_{13}$ is large enough, each degeneracy
parameter becomes same before BBN, namely $\xi_e =
\xi_{\mu,\tau}$. However, as recently stressed by Ref. \cite{PPR}, while this
conclusion is basically correct, oscillations take place so late that
perfect equilibrium is never achieved and the ratio of imperfectness
strongly depends on the magnitude of $\theta_{13}$ \cite{PPR}. It is
also pointed out that if a Nambu-Goldstone particle such as majoron on
interacts with neutrinos, flavor mixing can be suppressed and leads
$\xi_e \neq \xi_{\mu,\tau}$ \cite{majoron}. Therefore, we also consider
the second case $\xi_e \neq \xi_{\mu,\tau}$.
The second case is also motivated
by the fact that $\xi_\mu$ and $\xi_\tau$ have the same effects both on
the BBN and CMB because we consider the degenerate masses, while $\xi_e$
does not.

This paper is organized as follows. In the next section, we summarize
effects of $m_\nu$ and $\xi_i$ on BBN and CMB. In Sec. \ref{seclike}, we give the cosmological model that we use in parameter searching of MCMC
analysis. In Sec. \ref{secres}, we present our limits from the observational data based on WMAP5 and explain the correlation between the bound of each
parameter. In final section \ref{secsum}, we provide the summary of this paper.
Throughout this paper, we set $c = \hbar = 1$.

\section{cosmological effects of $\xi_i$ and $m_\nu$}\label{seceff}

In this section, we review the effects of the neutrino mass $m_\nu$ and
the degeneracy parameter $\xi_i$ on the predicted helium abundance $Y_p$
in BBN theory and the predicted CMB power spectrum $C_l$.
The degeneracy parameter of $i$th flavor of neutrino is defined as
\begin{eqnarray}
\xi_i \equiv \frac{\mu_{\nu_i}}{T_{\nu}},
\end{eqnarray}
where $\mu_{\nu_i}$ is the chemical potential.

\subsection{Effects on BBN} \label{seceffbbn}

The observed primordial light-element abundances constrain the
conditions during the BBN epoch from the time of weak reaction freeze
out ($t \sim 1 \rm sec$, $T \sim 1 {\rm MeV}$) to the freeze out of
nuclear reactions ($t \sim 10^4 \rm sec$, $T \sim 10 \rm {keV}$). In the
standard BBN theory, the abundance of light elements is determined by
only one parameter, namely, baryon-to-photon ratio. However, the lepton
asymmetry changes the abundance in two ways. 

One is that the extra
energy density coming from the neutrino degeneracy changes the expansion
rate of the universe during 
the BBN epoch and hence the primordial
abundances. Specifically, the Fermi-Dirac distribution functions of
neutrinos and anti-neutrinos 
with neutrino degeneracy are given by
\begin{eqnarray}
f_{\nu_i}(q,{\xi_i}) = \frac{1}{1 + {\rm e}^{q - \xi_i}}, \ \ \ 
f_{{\bar \nu}_i}(q,{\xi_i}) = \frac{1}{1 + {\rm e}^{q + \xi_i}},
\end{eqnarray}
where $q$ is comoving momentum written by proper momentum $P$ and
neutrino temperature $T_\nu$ as $q \equiv P/T_\nu$, and $\xi_i$ is the
degeneracy parameter of $i$th flavor of neutrino. The energy density for
massless degenerate neutrinos is given by 
\begin{equation}
\rho_{\nu} + \rho_{\bar \nu}
= \sum_i (k_B T_\nu)^4 \int \frac{d^3q}{(2 \pi)^3} q (f_{\nu_i}({q},\xi_i) +
f_{{\bar \nu}_i}({q},\xi_i)) 
\equiv \frac{N_{\rm eff}}{3} (\rho_{\nu_0} + \rho_{{\bar
\nu}_0}) ~,
\end{equation}
where $\rho_{\nu_0}$ is the energy density for massless non-degenerate
neutrinos and $N_{\rm eff}$ is the effective number of 
relativistic degrees of freedom of neutrinos given by
\begin{eqnarray}
  N_{\rm eff} &\equiv& \left( 1 + \frac{30}{7} \left( \frac{\xi_e}{\pi} \right)^2 + \frac{15}{7} \left( \frac{\xi_e}{\pi} \right)^4 \right) + 2 \left( 1 + \frac{30}{7} \left( \frac{\xi_{\mu,\tau}}{\pi} \right)^2 + \frac{15}{7} \left( \frac{\xi_{\mu,\tau}}{\pi} \right)^4 \right). \label{Neff} 
\end{eqnarray} 
In the non-degenerate neutrino case, $N_{\rm eff} = 3$. Therefore the
neutrino degeneracy increases the energy density of neutrino.  

The other is that positive $\xi_e$ suppresses $p \rightarrow n$ weak reactions and leads the neutron-to-proton ratio to smaller value both in and out of
the chemical equilibrium. Therefore the abundance of helium $Y_p$
decreases monotonically with increasing $\xi_e$. Note that the
dependence on $m_\nu$ can be neglected because $m_\nu$ is much smaller
than the temperature at the BBN epoch and does not affect 
the reaction rate.

\begin{figure}[t]
  \centering \includegraphics[width=8cm,height=6cm,clip]{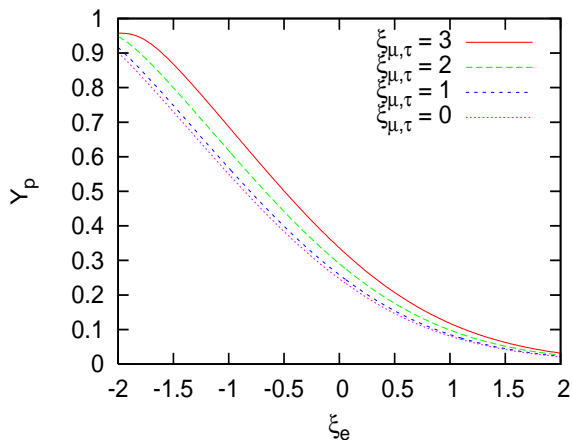}
  \caption{\label{figYp}
    Effects of the degeneracy parameters, $\xi_e$ and
    $\xi_{\mu,\tau}$, on the helium abundance $Y_p$. Here we fix the
    baryon abundance $\omega_b$ to the mean value from WMAP5 alone
    analysis for a power-law flat $\rm \Lambda$CDM model \cite{komatsu}.}
\end{figure}

We used the Kawano BBN code \cite{kawano} to include the two effects
described above and to calculate the abundance of light elements
correctly. We also updated the nuclear reaction rates \cite{rr}.  In
Fig. \ref{figYp}, the effects of $\xi_e$ and $\xi_{\mu,\tau}$ on the
helium abundance $Y_p$ are demonstrated. The larger $N_{\rm eff}$ makes
the universe expand faster and more neutrons survive before the
nucleosynthesis begins. Therefore the helium abundance becomes larger as
$\xi_{\mu,\tau}$ increases. On the other hand, $Y_p$ decreases
monotonically as $\xi_e$ increases because the chemical potential of
electron type neutrinos suppresses the neutron-proton ratio at the onset
of BBN as mentioned above.

\subsection{Effects on CMB}

Cosmological Microwave Background (CMB) photons are the remnants from the
Big-Bang in the early universe. They have evolved through the
interaction with the matters and the gravitational potentials, and have
been released from the last scattering surface at cosmological
recombination ($z \sim 1088$) to reach us through the large scale
structure of the universe.  Therefore CMB contains plenty of information
about the energy contents of the universe.  In the simplest cosmological
model, the universe is assumed to be spatially flat, and neutrinos are
massless and not degenerate.  The angular power spectrum of the CMB,
$C_l$, can be described by the six cosmological parameters, namely, the
energy density of baryon and dark matter, the present Hubble parameter,
the amplitude of the primordial curvature perturbation and its spectral
index, and the optical depth due to re-ionization.  However, if $m_\nu$
and $\xi_i$ are finite values, both of them also affect $C_l$ both at
the background and the perturbation levels.

In the background evolution, the mass of neutrino $m_\nu$ and the
neutrino degeneracy $\xi_i$ determine the energy density and pressure of
neutrinos. The energy density of massive degenerate neutrinos is
described as $\rho_\nu + \rho_{\bar \nu} = \sum_i m_\nu
(n_{\nu_i}(\xi_i) + n_{{\bar \nu}_i}(\xi_i))$, where ${n_{\nu_i}}$ is
the number density of neutrinos. Therefore the energy density parameter
of them, which is the normalized energy density by the critical density
$\rho_{\rm crit}= (3H^2)/(8 \pi G)$, is given as
\begin{eqnarray}
  \Omega_\nu &=& \frac {m_\nu (\Theta_2(\xi_e) +
  2\Theta_2(\xi_{\mu,\tau}))}{k_B T_\nu(0)} 
  \frac{\rho_{\nu_0}(0) + \rho_{\bar{\nu}_0}(0)}{\rho_{\rm crit}(0)}. \label{mnu}
\end{eqnarray} 
Here the argument $(0)$ means the value at present and the $n$th moment
is defined as
\begin{eqnarray}
  \Theta_n(\xi_i) \equiv \frac{1}{2} \int q^n dq (f_{\nu_i} + f_{\bar \nu_i})/ \left( \frac{7 \pi^4}{120} \right).
\end{eqnarray} 
When  we can regard neutrinos as massless degenerate particles, $\rho_\nu + \rho_{\bar \nu}$ is expressed by using $N_{\rm eff}$ just like the discussion in the previous subsection. 

Meanwhile, in the perturbation evolution, the $\xi_i$ changes on the
gravitational source term of the collisionless Boltzmann equation of
neutrino. Specifically, its coefficient, $df/dq$, is modified as
\begin{eqnarray}
  \frac{d {\rm ln} (f_{\nu_i} + f_{\bar \nu_i})}{d {\rm ln} q}
  = - \frac{q}{e^{-q} + {\rm cosh} \xi_i}
  \frac{1 + {\rm cosh}\xi_i{\rm cosh}q}{{\rm cosh}\xi_i + {\rm cosh}q}. \label{grav}
\end{eqnarray} 
Because we consider the situation that neutrinos have degenerate
masses, and because neutrinos are collisionless at the CMB epoch
irrelevant to their flavors, there are no difference in the effects on
$C_l$ between $\xi_e$ and $\xi_{\mu,\tau}$ contrary to the product
$Y_p$ at BBN. For the details of the effects of $m_\nu$ and $\xi$ on
CMB we refer readers to the references \cite{ichiki,pastor}. Here we
derived above equations following their notations.  
This dependence of the perturbation on $\xi$ is absent for the massless
  degenerate case, because the momentum dependence in distribution
  function is integrated away into the energy density.  In this case,
  the effect of $\xi_e$ on $C_l$ is identical to that of $N_{\rm
    eff}$ through Eq. (\ref{Neff}). Although this momentum dependence
  should have some distinctive effects on the evolution of
  perturbations in the massive degenerate case, we see that the dominant effect
  still comes from the change in $N_{\rm eff}$ for the neutrino masses
  $\sum m_\nu <1.5$ eV considered in this paper.

\begin{figure}[t]
  \begin{tabular}{cc}
    \begin{minipage}{0.5\hsize}
      \centering
      \includegraphics[width=8cm,height=6cm,clip]{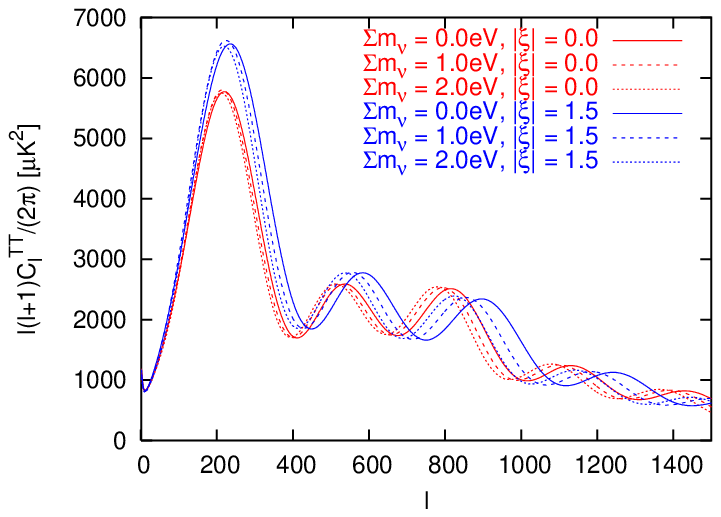}
      \caption{\label{figcmbxim}
        Effects of $m_\nu$ and $\xi$ on $C_l^{TT}$ 
        for the case $\xi_e = \xi_{\mu,\tau}$ with $Y_p$ 
        fixed as $0.24$. Here we fix
        $\omega_b + \omega_c$ and change $\omega_\nu$. Other cosmological
        parameters are taken as the mean value from WMAP5 alone analysis
        for a power-law flat $\rm \Lambda CDM$ model \cite{komatsu}
        i.e. $\Omega_b + \Omega_c + \Omega_\nu + \Omega_\Lambda = 1$.}
    \end{minipage}
    \begin{minipage}{0.5\hsize}
      \centering
      \includegraphics[width=8cm,height=6cm,clip]{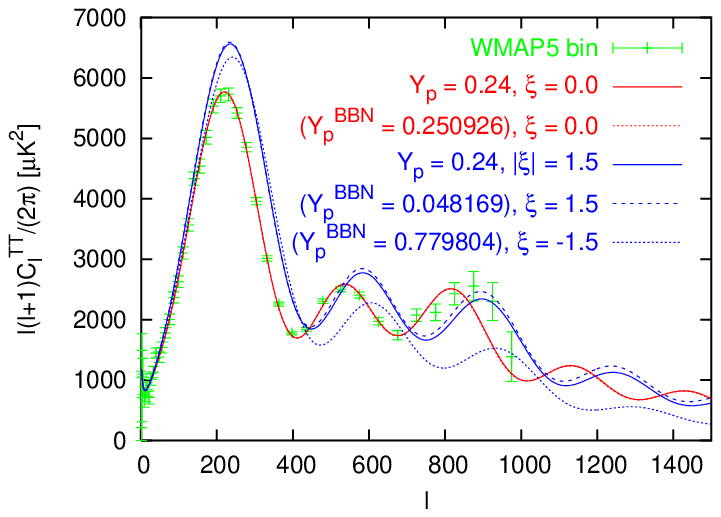}
      \caption{\label{figcmbxiYp}Effects of $\xi$ on $C_l^{TT}$ for the case $\xi_e = \xi_{\mu,\tau}$ with massless neutrinos. Two dotted lines and a dashed line are $C_l^{TT}$'s affected by $Y_p$ related with $\xi$ at BBN. Solid lines show $C_l^{TT}$'s affected by $Y_p$ unrelated with $\xi$. Here we fix $\omega_b + \omega_c$. Other cosmological parameters are taken as the mean value from WMAP5 alone analysis for a power-law flat
        $\rm \Lambda CDM$ model \cite{komatsu} i.e. $\Omega_b + \Omega_c +
        \Omega_\Lambda = 1$.}
    \end{minipage}
  \end{tabular}
\end{figure}

We inserted these two modifications in the Boltzmann Code for
Anisotropies in the Microwave Background (CAMB)
\cite{camb1,camb2,camb3}. Then we checked the consistency between the
$C_l$'s given by \cite{pastor} and those calculated using our modified
code. In Fig. \ref{figcmbxim}, effects of $m_\nu$ and $\xi$ on the CMB
temperature power spectrum $C_l^{TT}$ are demonstrated.

Now we explain the dependence of the CMB
spectrum on $\xi_i$ and $m_\nu$. If the effective number of neutrino, $N_{\rm eff}$, increases by
increasing $|\xi_i|$, the sound horizon at the last scattering becomes
smaller and the epoch of radiation-matter equality delays leading the
larger decay of the gravitational potential. These effects shift $C_l$ to
the higher $l$ and derive the enhancement of the 1st peak. In addition,
through the effects of free streaming of these ultra-relativistic
neutrinos, the smaller scale gravitational potentials have been damped
just prior to the beginning of the acoustic oscillations of the
baryon-photon fluid. This causes the smaller temperature fluctuations.
These effects on $C_l$ from $N_{\rm eff}$ are discussed in detail in
\cite{Neff}.


Next we focus our attention on the dependence on $m_\nu$
\cite{mnu,2008JPhCS.120b2004I}. In massive non-degenerate case ($\xi_i =
0$), neutrino mass of $O({\rm eV})$ causes overall horizontal shift and
suppression around the first peak (when we fix $\omega_b + \omega_c$ and
change $\omega_\nu$ to keep flatness).  The horizontal shift comes from
the fact that the larger $m_\nu$ (more non-relativistic particles at
present epoch) implies that the distance to the last scattering surface
is shorter and the peaks move to smaller $l$. However, this shift is
mostly canceled by the downward shift in $H_0$. Therefore this does not
produce a neutrino mass signal. If $m_\nu \gtrsim 0.6$\,eV, massive
neutrinos on average become non-relativistic before the epoch of
recombination and only in this case, the neutrino mass can significantly
imprint a characteristic signal in acoustic peaks (specifically, the
matter-radiation equality occurs earlier due to less relativistic
degrees of freedom and 1st peak is suppressed by the smaller
early-integrated Sachs-Wolfe effect) \cite{mnu,2008JPhCS.120b2004I}.

In massive degenerate case ($\xi_i \neq 0$), the behavior of $C_l$ can
be understood by combining the effects in the massless degenerate case
and the massive non-degenerate case.  In Fig. \ref{figcmbxim} we depict
the CMB angular power spectrum to illustrate the effects of mass and
degeneracy of neutrinos. We see that the $m_\nu$-dependence does not
much depend on $\xi$. As mentioned above, roughly speaking, neutrino
mass constraint from (WMAP-level) CMB observation is determined by a
critical mass $3m_{\nu,c}$ which is defined by $z_{rec} \simeq z_{nr}$,
where $z_{nr}$ denotes the redshift when neutrinos on average become
non-relativistic.  When there is a finite $\xi$, the neutrino momentum
distribution is modified and so is this critical mass. In this case, the
critical mass is obtained by
\begin{eqnarray}
\frac{m_{\nu,c}(\xi)}{m_{\nu,c}(\xi=0)} \simeq \frac{P_\nu (\xi)} {P_\nu (\xi = 0)}\simeq F_\nu(\xi),
\end{eqnarray}
where $P_\nu(\xi)$ and $P_\nu(\xi=0)$ are the average momenta of
neutrinos in massive degenerate and massive non-degenerate cases, given
by $P_\nu \approx \rho_\nu / n_\nu$, and $F_\nu$ is given by
\begin{eqnarray}
  F_\nu(\xi) & \equiv &
 \left(1 + \frac{30}{7}\left(\frac{\xi}{\pi}\right)^2 + \frac{15}{7}\left(\frac{\xi}{\pi}\right)^4\right) \frac{\Theta_2(\xi = 0)}{\Theta_2(\xi)}.
\end{eqnarray}

\begin{figure}[t]
   \centering
   \includegraphics[width=8cm,height=6cm,clip]{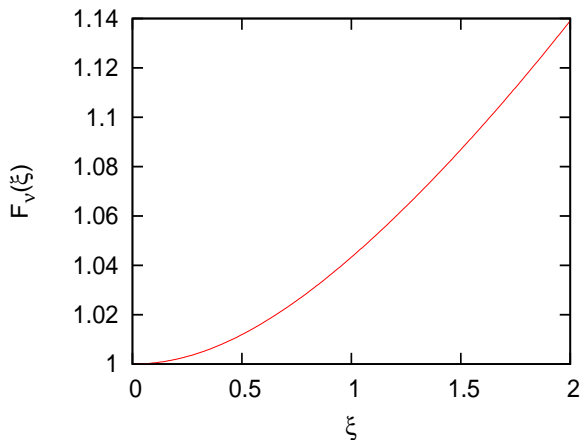}
   \caption{\label{figmc}
     The ratio of the neutrino masses between degenerate and non-degenerate cases for which the neutrinos become non-relativistic at recombination.}
 \end{figure}

Numerically we depict $F_\nu$ in Fig. \ref{figmc}. From the figure we
found $F_\nu \lesssim 1.14$ for $\xi < 2$. Therefore the critical mass
($3 m_{\nu, c}(\xi = 0) = 1.7 \ {\rm eV}$ for massive non-degenerate
case) shifts to, at most, $1.9 \ {\rm eV}$ for massive degenerate
case. As it is indicated \cite{mnu} that $m_\nu$ is already limited less
than $m_{\nu, c}$, we can expect that the upper bound of $\sum m_\nu$
from CMB is almost invariant even in the lepton asymmetric universe, as
will be shown later.

 Finally, we briefly explain the dependence of the CMB power spectrum
 on $Y_p$.  The main effect of $Y_p$ on the CMB power spectrum comes
 from the diffusion damping at small scales. When $Y_p$ is larger, the
 number of free electron becomes smaller, because the maximum number
 of free electron in the fully ionized universe is propotional to
 $(1-Y_p/2)$. The number of free electron becomes even smaller at the
 epoch of hydrogen recombination, since it is easier for electrons to
 recombine with $^4$He than with H and consequently more electrons
 have been captured into the $^4$He nuclei by that epoch.
 Thus the Compton mean free path becomes larger for larger $Y_p$,
 which mean that the diffusion length of photon becomes also
 larger. Since the photon-baryon tight coupling breaks down at the
 photon diffusion scales, the fluctuation of photon is exponentially
 damped due to the diffusive mixing and re-scattering. Hence the CMB
 power spectrum is damped more for larger values of $Y_p$. Detailed
 explanations are found in
 Refs.~\cite{2004PhRvD..69b3509T,2006PhRvD..73f3528I,Yp}. This effect
 is rather small and not relevant to WMAP5 level observation. However,
 in our case, since $Y_p$ can take very large or small value with $\xi
 \sim O(1)$, this may affect constraints.  For example, two models
 with $\xi_e=\xi_{\mu,\tau}=1.5$ and $-1.5$ give the same CMB spectra
 if we fix $Y_p=0.24$. However, if we consider the effects of $\xi$ at
 BBN epoch correctly, the two models predict $Y_p=0.048$ and $0.780$,
 respectively as shown in Fig. \ref{figYp}. Therefore, the model with
 negative degeneracy parameter should lead the larger diffusion
 damping as shown in Fig. \ref{figcmbxiYp}.  The CMB power spectrum
 can be affected by $\xi$ through such an indirect manner
 \cite{2008JCAP...03..004H}.
  
 \section{Model and likelihood analysis}\label{seclike}

In what follows we will put constraints on the neutrino degeneracy
parameters $\xi_i$ and masses $m_\nu$ along with the other standard
cosmological parameters. For this purpose we implement CAMB and BBN
codes modified by us for lepton asymmetric cosmology in the CosmoMC
Monte Carlo Markov Chain (MCMC) public package \cite{cosmomc}. In our
analysis we use observational data set in several combinations; WMAP5
alone, WMAP5 + prior on Hubble parameter ($H_0=72 \pm 8$ km/sec/Mpc
(68\%C.L.)) given by Hubble Space Telescope (HST) \cite{HST}, and WMAP5 + $^4$He and
D/H abundances (BBN). Throughout our analysis we assume that the
universe is spatially flat and neutrinos are degenerate in mass. We
explore the likelihood in 10 
dimensional parameter space, i.e.,
\begin{eqnarray}
  \vec{p} = (\omega_b, \omega_{dm}, A_s, n_s, \theta_s, \tau, A_{\rm
   SZ}, \xi_e, \xi_{\mu,\tau}, \omega_\nu). 
\end{eqnarray}
Here $\omega_b$, $\omega_{\rm dm}$ and $\omega_\nu$ are the energy
densities of baryon, dark matter, and massive neutrino, respectively,
$A_s$ and $n_s$ are the amplitude and the spectral index of the
primordial density perturbation power spectrum at the pivot
scale $k_* = 0.002$ Mpc$^{-1}$, $\theta_s$ is the acoustic peak scale,
$\tau$ is the optical depth of reionization, and $A_{\rm SZ}$ is the
amplitude parameter of thermal Sunyaev-Zel'dovich (SZ) effect \cite{SZ}.
Note that $\omega_{\nu}$ is included in $\omega_{dm}$ so that
$\omega_{dm} \equiv \omega_{c} + \omega_{\nu}$, where $\omega_c$ is the
energy density of cold dark matter. 
To calculate $C_l$ in the parameter estimation, we estimated $H_0$ from $\theta_s$ by using the fitting fomula in $\Lambda$CDM for the redshift of recombination $z_{\rm rec}$. This is the default method in CosmoMC. In the degenerate case, we should consider the effect of $\xi$ on this fitting formula. However, as we found that the shift of $z_{\rm rec}$ for $\xi = 2$ is less than 0.1\% from the result of the fitting fomula in $\Lambda$CDM for $\xi = 0$, we used it safely even in the degenerate case.
As for the helium abundance $Y_p$
 used in CMB calculations, we consider two situations: one is
that we (approximately) fix $Y_p=0.24$ regardless of the cosmological
parameters, and the other is that $Y_p$ is derived at each time of MCMC
step from the other cosmological parameters using the modified
BBN code. Hereafter we denote for the latter case the helium abundance
as $Y^{\rm BBN}_p$, i.e.,
\begin{eqnarray}
Y_p = Y_p^{\rm BBN}(\omega_b, \xi_e, \xi_{\mu,\tau})~.
\label{Eq:YpBBN}
\end{eqnarray}
Note that for the former case we vary the degeneracy parameter $\xi$
freely in the MCMC analysis even if the helium abundance is kept
fixed as $Y_p=0.24$.

In the usual CMB constraints the helium abundance is often fixed to
the standard value, say $Y_p=0.24$.  This condition is justified by
the fact that current CMB measurements have already put a tight
constraint on $\omega_b$, and for this constrained range of $\omega_b$
the CMB spectrum does not change very much even if we allow the helium
abandance to vary according to the standard BBN prediction.

However, in the lepton asymmetric universe considered in this
paper, the value of $Y_p$ can deviate significantly from $Y_p=0.24$
depending on $\xi$ as discussed above. In this sense, any analysis of
CMB constraint on $\xi$ with fixed $Y_p$ is inconsistent unless $\xi$
is constrained tightly.  In our CMB power spectrum calculation as is
used in later analysis, $Y_p$ is set by the BBN theoretical
calculation as a function of $\omega_b$, $\xi_e$ and $\xi_{\mu,\tau}$.
We also investigate constraints in the lepton asymmetric models with
fixed $Y_p$ for comparison.

\section{Results and comparisons}\label{secres}

In this section, we describe our results and give their interpretations.
For a test calculation we first ran CosmoMC to constrain neutrino masses
from WMAP5 alone with $\xi=0$. We obtained consistent but slightly
tighter result, $\sum m_\nu < 1.2 \ \rm eV \ (\rm 95 \% CL)$, while WMAP
team obtained $1.3$ eV \cite{dunkley}. This difference would be
attributed to differences in the treatment of cosmological parameters as
well as their priors \cite{2009PhRvD..79b3520I} \cite{cmbws}.  In
general, posterior distributions depend on prior distributions in
Bayesian estimation \cite{cosmomc}. 

Next, we ran a MCMC with all of the parameters to be fixed other than
$\xi_{\mu, \tau}$ and confirmed that the 1D posterior distributions of
$\xi_{\mu, \tau}$ are symmetric about $\xi=0$. Using this fact we
restricted the prior range of $\xi_{\mu,\tau}$ only to positive values,
i.e., we actually estimate $|\xi_{\mu,\tau}|$ from MCMC analysis.

We perform MCMC analyses for three combinations of observational data
sets described below.  In all analyses, we neglect effects of CMB
lensing because they slightly change $C_l$ but do not change the
posterior distributions very much.

\subsection{Constraints from WMAP5 alone}

We first present the result of constraints on cosmological parameters
from WMAP5 alone in TABLE \ref{wmap}.  In that table, we focus on 5
parameters, $\xi_i, m_\nu, H_0, Y_p, \omega_{dm}$. For cosmological
models we consider the following four cases and give the results 
separately.  We present the mean values and 68\% and 95\% confidence
regions for the case $\xi_e=\xi_{\mu,\tau}$ in the left panel of TABLE
\ref{wmap}, and for the case $\xi_e \neq \xi_{\mu,\tau}$ in the right
panel. For each 
panel we separate the results for cases whether we use BBN code to
calculate $Y_p$ or we approximately fixed $Y_p$ value to $0.24$. 
The results for the massless neutrino cases are given in the appendix.

As is found in TABLE \ref{wmap}, the constraints on $\xi$ from CMB are
highly dependent on whether one fixes helium abundance or derives it
from BBN theory. 
First, let us consider the case with fixed helium abundance. In this
case, the constraints on $\xi$ from CMB becomes symmetric about
$\xi_i=0$, because positive and negative $\xi$ give identical effects on
the CMB spectrum through Eqs. (\ref{Neff}) (\ref{mnu}), and (\ref{grav}). 
In reality, however, the
positive and negative $\xi$ give different effects on the CMB spectrum
through the helium abundance.
In CMB theory, helium abundance is related to the
number density of free electrons, and hence the Silk damping scale
depends on it. So the CMB can put constraint on $Y_p$, and therefore
$\xi$ is further constrained through this effect if one derives 
helium abundance from BBN theory for the CMB calculations.  From the
fact that the current 
upper limit on helium abundance from CMB is found to be $Y_p< 0.44$ and
there is no lower bound \cite{Yp}, negative $\xi_e$ is disfavored,
because negative $\xi_e$ leads to large $Y_p$ as is discussed in \ref{seceffbbn}.  
As for the other cosmological parameters,
such as $\omega_b$ and $n_s$, are tightly constrained for all cases and
the constraints do not change between different models considered
here, because the dependence of $C_l$ on these parameters does not
degenerate with that on $\xi_i$ and $Y_p$. 

\begin{figure}[t]
  \begin{tabular}{cc}
    \begin{minipage}{0.5\hsize}
      \centering
      \includegraphics[height=5.7cm,clip]{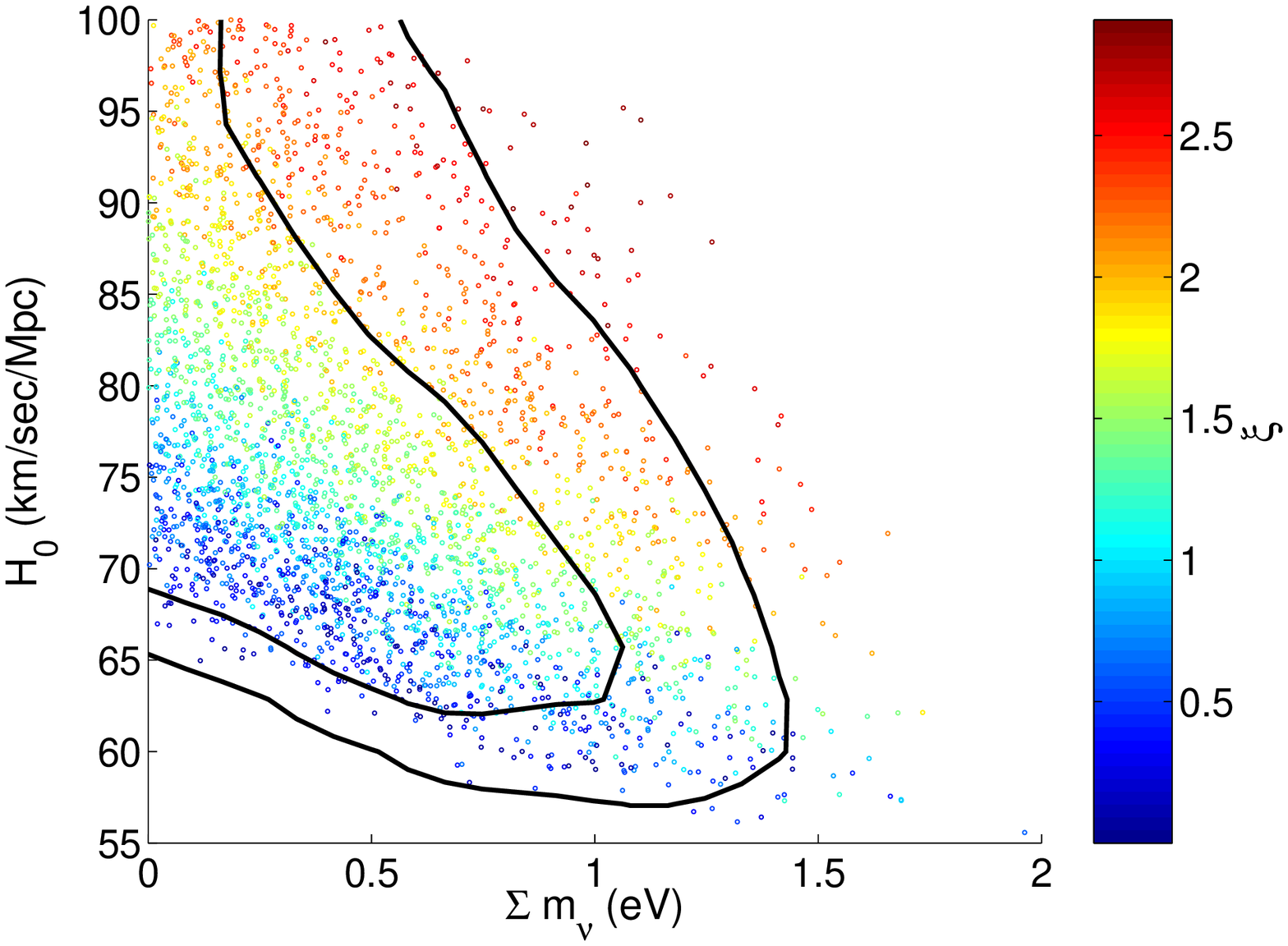}
    \end{minipage}
    \begin{minipage}{0.5\hsize}
      \centering
      \includegraphics[height=5.7cm,clip]{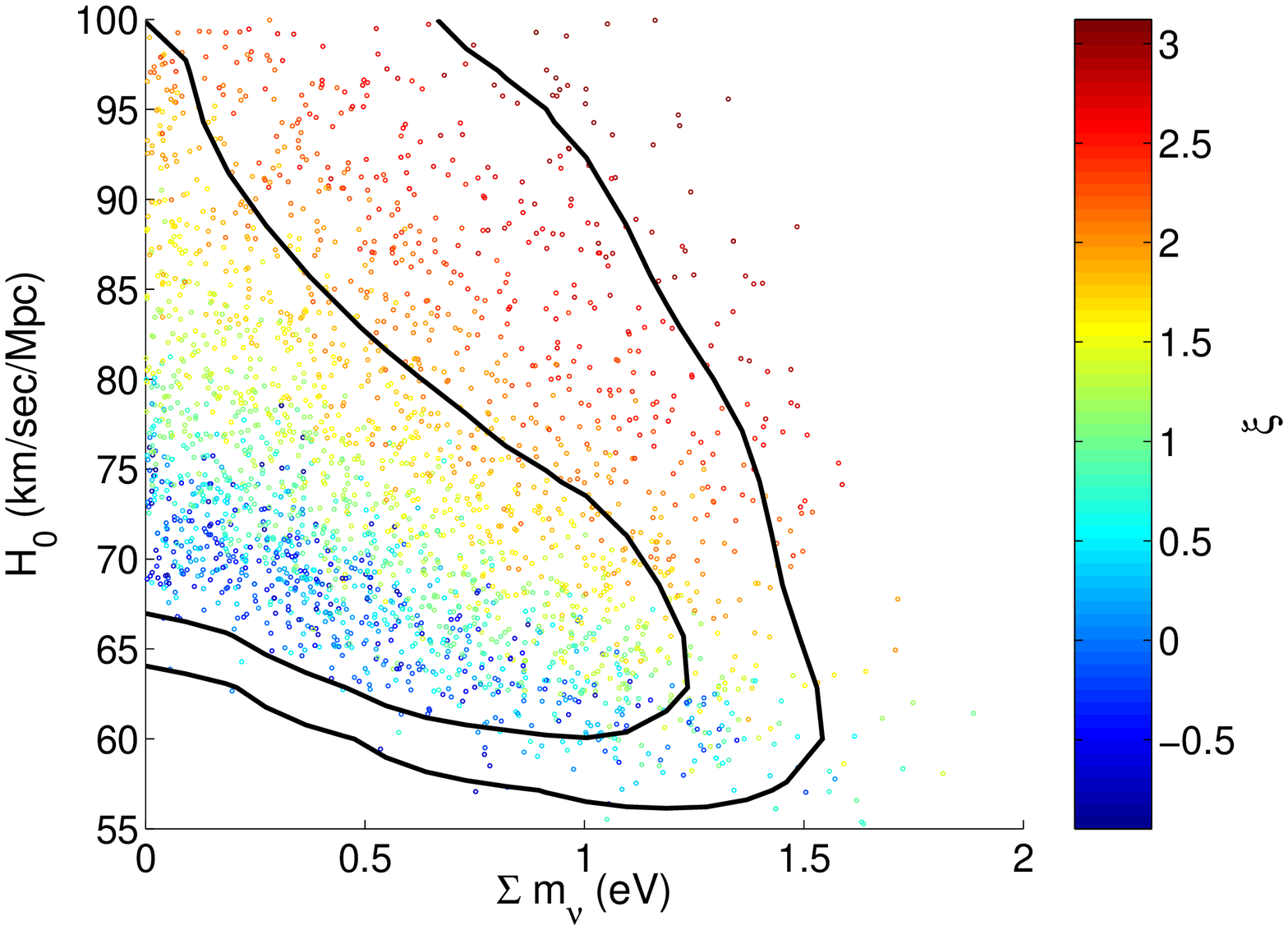}
    \end{minipage}
  \end{tabular}
  \caption{\label{fig3Dwmap}
    The constraints in the 3D region of $\sum m_\nu$, $H_0$
    and $\xi$ from WMAP5 alone in the case of $\xi_e = \xi_{\mu,\tau}$
    with $Y_p = 0.24$ (the left figure) or $Y_p = Y_p^{\rm BBN}$ (the right
    one). The density of the points, whose colors are coded according to
 the value of $\xi$, represent the probability 
 distribution from MCMC samples. The contours are the 68\% and 95\%
 confidence regions. }
\end{figure}

One may find it inconsistent that the constraints on $\xi_e$ and
$\xi_{\mu,\tau}$ with fixed $Y_p = 0.24$ are different because in this
case $\xi_e$ and $\xi_{\mu,\tau}$ play the same role in the
CMB. However, in fact, the bound of $\xi_{\mu,\tau}$ should be tighter
than that of $\xi_e$ because we implicitly assume the condition
$\xi_\mu = \xi_\tau$ in this analysis. 
We analytically
estimated this difference as $|\xi_e/\xi_{\mu,\tau}| \sim 1.3$, which can
be derived as follows. If we varied each of the degeneracy parameter
freely, the distributions for $\xi_e$, $\xi_\mu$ and $\xi_\tau$ should
 have been identical in the case where helium abundance is kept fixed and the
masses are degenerated. The effects of $\xi$ in this case mainly come
through an increase of neutrino energy density, i.e.,
Eq. (\ref{Neff}). Therefore, once the constraints on $\xi_e$ can be
obtained in the form as
\begin{equation}
\frac{30}{7}\left(\frac{\xi_e}{\pi}\right)^2+\frac{15}{7}\left(\frac{\xi_e}{\pi}\right)^4
 < \alpha~,
\end{equation}
with some value of $\alpha$, then, the constraints on $\xi_{\mu,\tau}$ should read
\begin{equation}
2\left[\frac{30}{7}\left(\frac{\xi_{\mu,\tau}}{\pi}\right)^2+\frac{15}{7}\left(\frac{\xi_{\mu,\tau}}{\pi}\right)^4\right]
 < \alpha~.
\end{equation}
We found
that this estimate is in good agreement with our MCMC constraints.
The result of the constraints from WMAP5 alone is depicted on
$H_0$-$\sum m_\nu$ plane in Fig. \ref{fig3Dwmap}. In that figure, the
points are drawn such that the density of the points is proportional to
the number of the MCMC samples (i.e., probability distributions), and
colors represent the value of $\xi$. One can see from this figure a
well-known $H_0$-$\sum m_\nu$ degeneracy with fixed $\xi$ values. 
We find that the upper bound of $\sum m_\nu$ is not changed
significantly because this upper bound is lower than the critical mass
$3m_{\nu,c}$.
It is
interesting 
to note that the massive neutrino cosmology with $\sum m_\nu \approx 1 $eV
is compatible with the Hubble parameter $H_0\approx 72$ km/sec/Mpc
if one allows $\xi\approx 1$ lepton asymmetry, as will be discussed in
the next subsection.

\begin{table}[t]

 \begin{center}
  \begin{tabular}{ccccc}
   \hline   
   & \multicolumn{2}{c}{$\xi \equiv \xi_e = \xi_{\mu, \tau}$} &
   \multicolumn{2}{c}{$\xi_e \neq \xi_{\mu, \tau}$} \\
   parameter & $Y_p = 0.24$ & $Y_p =  Y_p^{\rm BBN}$ & $Y_p = 0.24$ & $Y_p =  Y_p^{\rm BBN}$ \\
   \hline \hline
   $100 \omega_b$ & $2.205 \pm 0.070$ & $2.206^{+0.072}_{-0.074}$ & $2.200^{+0.069}_{-0.070}$ & $2.199 \pm 0.075$ \\
   $\omega_{dm}$ & $0.1823^{+0.015}_{-0.043}$ & $0.1799^{+0.0079}_{-0.0501}$ &
   $0.2111^{+0.0202}_{-0.0477}$ & $0.2158^{+0.0208}_{-0.0575}$ \\
   $\xi_e$ & $< 2.41$ & $1.20^{+0.50}_{-0.40}{}^{+1.39}_{-1.61}$ &
   $< 3.47$ & $1.65^{+0.63}_{-0.68}{}^{+2.00}_{-1.84}$ \\
   $|\xi_{\mu,\tau}|$ & - & - & $< 2.70$ & $< 2.81$ \\
   \hline
   $\sum m_\nu$ (eV) & $< 1.2$ & $< 1.3$ & $< 1.2$ & $< 1.3$ \\
   $Y_p$ & - & $< 0.36$ & - & $< 0.32$ \\
   \hline
 \end{tabular}
  \caption{Mean values and 68\% C.L. errors of the cosmological
  parameters obtained from the analysis of WMAP5 alone in the lepton
  asymmetric universe with massive neutrinos. For the constraints on $\xi$, 95\%
  C.L. errors are also shown, and the upper bounds on $\xi$, $\sum m_\nu$,
  and $Y_p$ are at 95\% C.L.
  The left side of the table
  is in the case of $\xi_e = \xi_{\mu, \tau}$. The right side is in the
  case $\xi_e \neq \xi_{\mu, \tau}$. Two parameters at the bottom are
  the derived parameters from the MCMC sampling.}  \label{wmap}
 \end{center}
\end{table}


From WMAP5 alone, the upper bounds of $\xi_i$ are actually determined by
the flat prior on the $H_0$ which is taken as $H_0 < 100$ km/sec/Mpc.
Therefore, analysis which takes into account external prior on $H_0$
should be performed. In the next subsection, we apply a prior of $H_0$
from HST.

\subsection{Constraints from WMAP5 + HST}
In this subsection, we present the constraints from WMAP5 + HST, with
the helium abundance being set to $Y_p^{\rm BBN}$ (see Eq. (\ref{Eq:YpBBN})). Here
the prior from HST is $H_0 = 72 \pm 8$ km/sec/Mpc (68\%C.L.) \cite{HST}. These
results are presented in TABLE \ref{wmaphst}. 
 
In TABLE \ref{wmaphst}, we found that all the parameters except for
$\sum m_\nu$ and $Y_p$ are limited more tightly than those limited from
WMAP5 alone. This is due to breaking the $H_0$ - $\xi$ degeneracy.
       
In the standard non-degenerate case, by adding HST prior the upper bound
of $\sum m_\nu$ becomes tighter than that from WMAP5 alone, and we found
$\sum m_\nu < 1.0$ eV.  This is because the prior of HST breaks
effectively the well-known $\sum m_\nu$ - $H_0$
degeneracy. Interestingly, however, we found that the upper bounds of
$\sum m_\nu$ from WMAP5 + HST do not change very much compared with the
upper bounds from WMAP5 alone if the lepton asymmetry is allowed. In the
degenerate 
case, higher neutrino masses are compatible with the Hubble parameter
around $H_0\approx 72$ km/sec/Mpc with non-zero $\xi$, as depicted in Fig. \ref{fig3Dwmaphst}. This is because the shift of the peak to the lower
$l$ due to larger $m_{\nu}$ can be compensated by the shift due to
larger $\xi$, with the Hubble parameter fixed. Therefore, HST prior does
not improve the limits on neutrino masses in the lepton asymmetric
universe.

\begin{table}[t]
  \begin{center}
    \begin{tabular}{ccc}
      \hline
      parameter & $\xi \equiv \xi_e = \xi_{\mu,\tau}$ & 
      $\xi_e \neq \xi_{\mu, \tau}$ \\
      \hline \hline
      $100 \omega_b$ & $2.202^{+0.069}_{-0.072}$ & $2.193^{+0.072}_{-0.073}$ \\
      $\omega_{dm}$ & $0.1621^{+0.0045}_{-0.0384}$ & $0.1947^{+0.0132}_{-0.0451}$ \\ 
      $\xi_e$ & $0.96^{+0.50}_{-0.40}{}^{+1.35}_{-1.52}$ & $1.48^{+0.55}_{-0.64}{}^{+1.92}_{-1.68}$ \\
      $|\xi_{\mu,\tau}|$ & - & $< 2.50$ \\
      \hline
      $\sum m_\nu$ (eV) & $< 1.3$ & $< 1.4$ \\
      $Y_p$ & $< 0.41$ & $< 0.32$ \\
      $H_0$ (km/sec/Mpc) & $71.8_{-3.0}^{+2.2}$ & $74.0_{-3.4}^{+2.8}$ \\
      \hline
    \end{tabular}
    \caption{\label{wmaphst}
      Mean values and 68\% C.L. errors of the
      cosmological parameters obtained from the analysis of WMAP5 + HST
      in the lepton asymmetric universe with massive neutrinos. For the
      constraints on $\xi$, 95\%
      C.L. errors are also shown, and the upper bounds on $\xi$, $\sum m_\nu$,
      and $Y_p$ are at 95\% C.L. The left
      side of the table is in the case of $\xi_e = \xi_{\mu, \tau}$. The
      right side is in the case of $\xi_e \neq \xi_{\mu, \tau}$.  Here
      we use the relation $Y_p = Y_p^{\rm BBN}$ in the calculation of the
      CMB power spectra. Three parameters at the bottom are the derived
      parameters from the MCMC sampling.}  
  \end{center}
    \end{table}
    
    
    \begin{figure}[t]
      \begin{tabular}{cc}
        \begin{minipage}{0.5\hsize}
          \centering
          \includegraphics[height=5.8cm,clip]{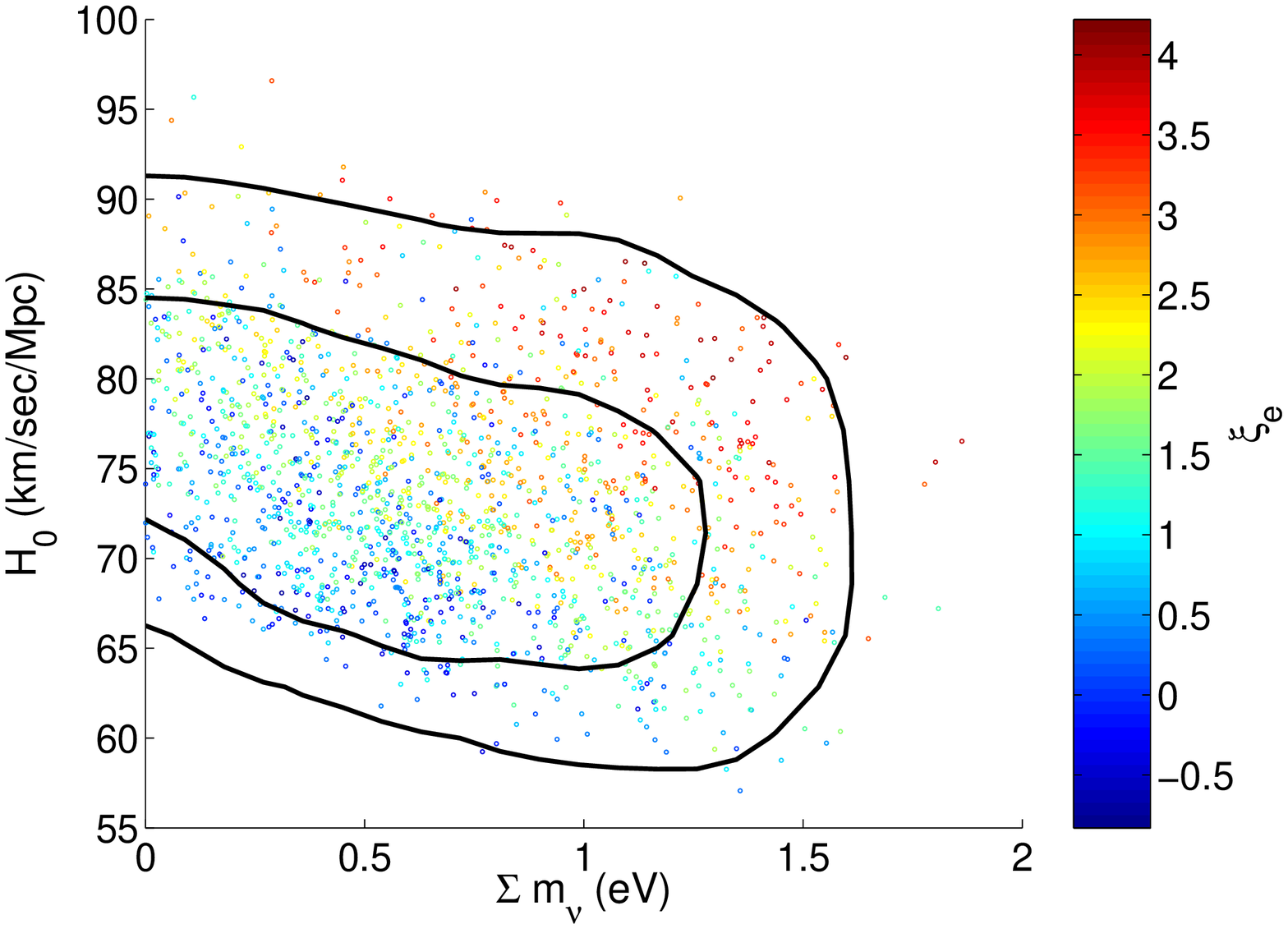}
        \end{minipage}
        \begin{minipage}{0.5\hsize}
          \centering
          \includegraphics[height=5.8cm,clip]{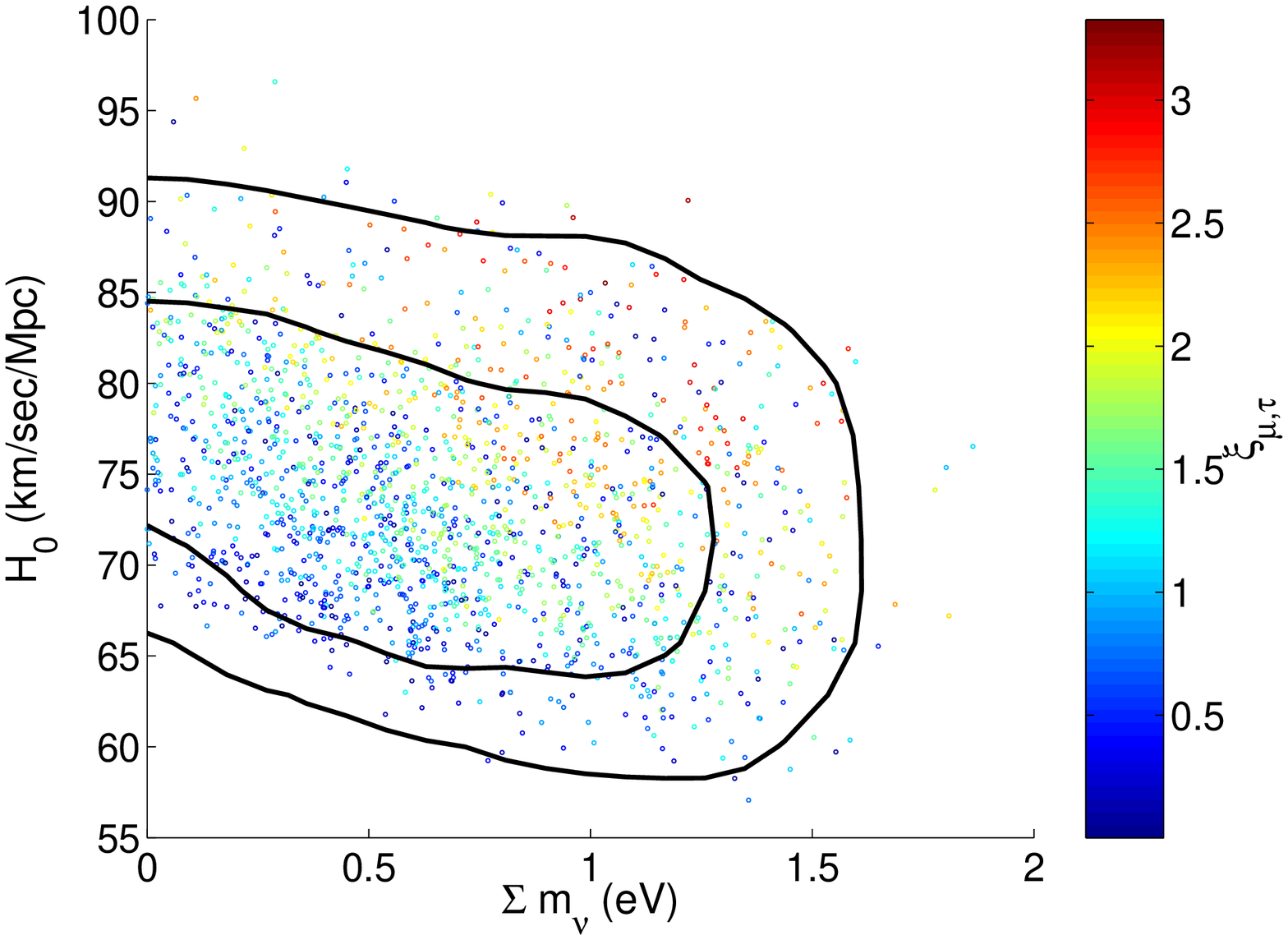}
        \end{minipage}
      \end{tabular}
  \caption{
    The constraint in the 3D region of $\sum m_\nu$, $H_0$
    and $\xi_e$ (the left figure) or $\xi_{\mu,\tau}$ (the right one) from
    WMAP5 + HST in the massive $\xi_e \neq \xi_{\mu,\tau}$ case with $Y_p =
    Y_p^{\rm BBN}$. The points and contours are depicted in the same way as
    Fig. 5.}
\end{figure} 

\subsection{Constraints from WMAP5 + BBN}   

Finally, we present the constraints from WMAP5 combined with
observations of light element abundances of $Y_p$ and D/H. 
We use only $Y_p = 0.250 \pm 0.004$ \cite{Ypobs} in the case $\xi_e = \xi_{\mu,\tau}$.  In
the case $\xi_e \neq \xi_{\mu, \tau}$, there is a parameter degeneracy
between $\xi_e$ and $\xi_{\mu,\tau}$ for the constraint from the helium
abundance alone. Therefore, we further use the deuterium abundance
constraint as $D/H = (2.82 \pm 0.27) \times 10^{-5}$ \cite{Dobs} in this
case to break the degeneracy.  The results are listed in
TABLE. \ref{wmapbbn}, and our constraints on the sum of neutrino
masses are 1.2\,eV and 1.1\,eV 
respectively for the cases $\xi_e = \xi_{\mu, \tau}$ and $\xi_e \neq \xi_{\mu, \tau}$. They do not depend on the assumption of
$\xi_e = \xi_{\mu, \tau}$ and $\xi_e \neq \xi_{\mu, \tau}$. Again, we
have found that neutrino mass constraints do not change much depending
on the assumptions on $\xi$.

As for the constraints on the degeneracy parameters, we obtain
$\xi = -0.013^{+0.016}_{-0.017}$ in the case $\xi_e = \xi_{\mu, \tau}$, and
$\xi_e = 0.034^{+0.017}_{-0.025}$ and $|\xi_{\mu,\tau}|<1.60$ in the case
$\xi_e \neq \xi_{\mu, \tau}$ from WMAP5 + BBN. In relation to them, we also performed
a conventional analysis. Namely, we searched three-dimensional parameter
space of  $\omega_b$, $\xi_e$ and $\xi_{\mu,\tau}$ by BBN calculation. Here, we take the baryon density of $\omega_b
= 0.0225 \pm 0.0006$ obtained from WMAP5 data in flat $\Lambda$CDM model.
The constraints, {$\xi = -0.013^{+0.016}_{-0.017}$ in the case $\xi_e = \xi_{\mu, \tau}$, and
$\xi_e = 0.002^{+0.056}_{-0.024}$ and $|\xi_{\mu,\tau}|<1.61$ in the case
$\xi_e \neq \xi_{\mu, \tau}$} are consistent with the cases above
confirming the validity of the conventional procedure. 

\begin{table}[t]
  \begin{center}
    \begin{tabular}{ccc}
      \hline
      parameter & $\xi \equiv \xi_e = \xi_{\mu,\tau}$ & 
      $\xi_e \neq \xi_{\mu, \tau}$ \\
      \hline \hline
      $100 \omega_b$ & $2.215 \pm 0.066$ & $2.221 \pm 0.061$ \\
      $\omega_{dm}$ & $0.1147^{+0.0079}_{-0.0078}$ & $0.1311^{+0.0056}_{-0.0096}$ \\ 
      $\xi_e$ & $-0.013^{+0.016}_{-0.017} \pm 0.033$ & $0.034^{+0.017}_{-0.025}{}^{+0.075}_{-0.058}$ \\
      $|\xi_{\mu,\tau}|$ & - & $< 1.60$ \\
      \hline
      $\sum m_\nu$ (eV) & $< 1.2$ & $< 1.1$ \\
      $Y_p$ & $0.250 \pm 0.004$ & $0.250 \pm 0.004$ \\
      $H_0$ (km/sec/Mpc) & $66.4^{+4.4}_{-4.6}$ & $70.0^{+2.7}_{-3.1}$ \\
      \hline
    \end{tabular}
    \caption{Mean values and 68\% C.L. errors of the
      cosmological parameters obtained from the analysis of WMAP5 + BBN
      in the lepton asymmetric universe with massive neutrinos. For the
   constraints on $\xi$, 95\%
  C.L. errors are also shown, and the upper bounds on $\xi$ and $\sum m_\nu$
  are at 95\% C.L.. The left
      side of the table is in the case of $\xi_e = \xi_{\mu, \tau}$. The
      right side is in the case of $\xi_e \neq \xi_{\mu, \tau}$. Here we
      use the relation $Y_p = Y_p^{\rm BBN}$ in the calculation of the CMB
      power spectra. Three parameters at the bottom are the derived
   parameters from the MCMC sampling.}  \label{wmapbbn}
  \end{center}
\end{table}
 
\section{Summary}\label{secsum}

\begin{figure}[t]
  \begin{tabular}{cc}
    \begin{minipage}{0.5\hsize}
      \centering
      \includegraphics[width=8cm,height=6cm,clip]{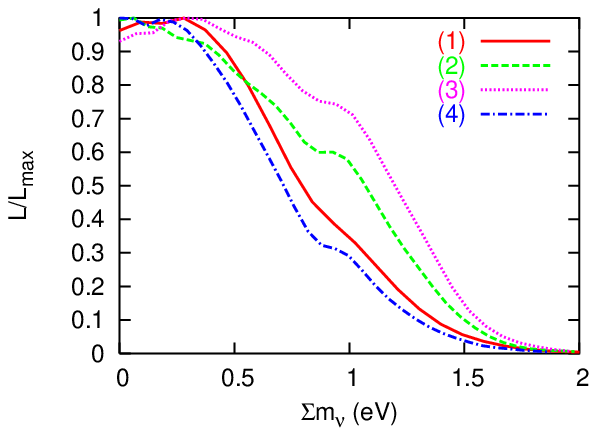}
    \end{minipage}
    \begin{minipage}{0.5\hsize}
      \centering
      \includegraphics[width=8cm,height=6cm,clip]{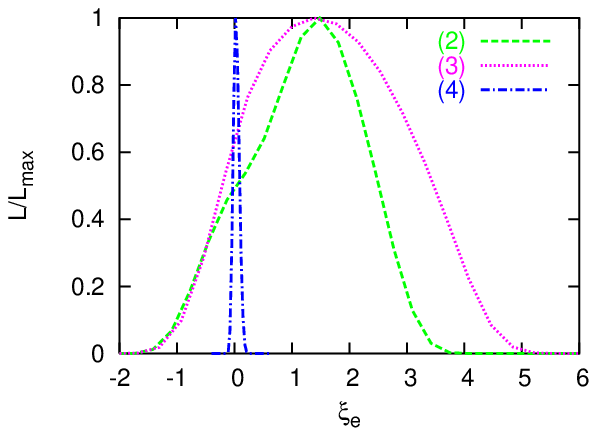}
    \end{minipage}
  \end{tabular}
  \caption{\label{fig1Dlike}
    One dimensional posterior distributions of $\sum m_\nu$ (the left panel) and $\xi_e$ (the right one) in the
    specific 4 cases. Case (1): the WMAP5 alone constraint
    with $Y_p=0.24$ and no lepton asymmetry (red solid line), Case (2): the WMAP5 alone constraint with $Y_p=Y_p^{\rm BBN}$ and $\xi_e=\xi_{\mu,\tau}$ (green dashed line), Case (3): WMAP5 alone constraint with $Y_p=Y_p^{\rm BBN}$ and
    $\xi_e\neq \xi_{\mu,\tau}$ (magenta dotted line), Case (4): the WMAP5+BBN constraint
    with $Y_p=Y_p^{\rm BBN}$ and $\xi_e\neq \xi_{\mu,\tau}$ (blue dot-dashed line).}
\end{figure}

In this paper, we have discussed the effects of neutrino masses and
neutrino degeneracies on CMB and BBN, and investigated how the latest
observations put constraints on these properties simultaneously. In
particular, we have examined how robust the constraint on neutrino
masses is in the lepton asymmetric universe. Our constraints on the
neutrino masses are $m_\nu<1.3$\,eV at 95\% C.L. from WMAP alone.
The constraint does
not change very much even if we put the HST prior in the lepton
asymmetric universe, as explained in
Sec. 4.2.  
We investigate the two types of lepton
asymmetry, i.e. the cases with $\xi_e = \xi_{\mu, \tau}$ and $\xi_e \neq
\xi_{\mu, \tau}$, and found that the constraints are almost the same for both
cases. We
also have taken into account the measurements of light element
abundances, and found that changes in constraints are very small.
These results are summarized in the left panel of Fig. \ref{fig1Dlike}.

As for constraints on the degeneracy parameters, we have obtained at
95\% C.L. $-0.41<\xi_e < 2.59$ for $\xi_e = \xi_{\mu, \tau}$. In the
case of $\xi_e \neq \xi_{\mu, \tau}$, the constraints are $-0.19<\xi_e <
3.65$ and $|\xi_{\mu, \tau}|<2.81$ from WMAP5 alone. We point out that
they are limits 
obtained by imposing the BBN relation for $Y_p$ and this helps to place
tight lower bound on $\xi_e$. 
A part of these results are shown in the right panel of Fig. \ref{fig1Dlike}. There, we can notice that the posteriors are skewed. In addition, the right panel indicates that $\xi_e$'s prefer the positive value, however, they are still consistent with $\xi_e = 0$.
Although these constraints are much weaker
than those obtained from $Y_p$ measurements in HII regions and direct
use of BBN theory, this would be more useful in future CMB measurements 
 such as Planck \cite{2008JCAP...03..004H, popa} to constrain $\xi_e$.


\begin{acknowledgments}
This work is supported in part by the Grant-in-Aid for Scientific
Research from the Ministry of Education, Science, Sports, and Culture of
Japan No.\,21740177 (K. Ichiki) and No.\,21740187 (M.Y.).
This work is supported in part by the Grant-in-Aid for Scientific
Research on Priority Areas No. 467 "Probing the Dark Energy through an
 Extremely Wide and Deep Survey with Subaru Telescope" and by the
 Grant-in-Aid for Nagoya University Global COE Program, "Quest for
 Fundamental Principles in the Universe: from Particles to the Solar
 System and the Cosmos", from the Ministry of Education, Culture,
 Sports, Science and Technology of Japan. 

\end{acknowledgments}

\appendix
\section{Constraints for the massless degenerate case}\label{secappen}

In this Appendix, we present the result of constraints on cosmological
parameters for massless degenerate case with $Y_p$'s relation as $Y_p =
Y_p^{\rm BBN}$. In TABLE \ref{xi1m0bbnYp} and \ref{xi2m0bbnYp}, we show all
the results. In this case, basically, the bounds are determined by 
the similar effects as we discussed in the massive case. Here we stress
the point simplified by dealing with neutrinos as massless particles is
that $H_0$ is completely determined by the degeneracy effect of $H_0 \ -
N_{\rm eff}$ reported in \cite{Neff} once $\xi$'s is limited by the effect of
$Y_p$.

\begin{table}[h]
  \begin{center}
    \begin{tabular}{cccc}
      \hline
      parameter & WMAP5 alone & WMAP5 + HST & WMAP5 + BBN \\
      \hline \hline
      $100 \omega_b$ & $2.258 \pm 0.063$ & $2.251^{+0.063}_{-0.062}$ &
      $2.250^{+0.063}_{-0.062}$ \\
      $\omega_{dm}$ & $0.1421^{+0.0079}_{-0.0254}$ & $0.1220^{+0.0031}_{-0.0106}$ &
      $0.1080 \pm 0.0062$ \\ 
      $\xi$ & $0.91^{+0.54}_{-0.41}{}^{+1.21}_{-1.50}$ & $0.42^{+0.39}_{-0.36}{}^{+1.05}_{-1.13}$ & $-0.013 \pm 0.017 {}^{+0.034}_{-0.032}$ \\   
      \hline
      $N_{\rm eff}$ & $< 10.2$ & $< 6.18$ & $< 3.002$ \\
      $Y_p$ & $< 0.42$ & $< 0.45$ & $0.250 \pm 0.004$ \\
      $H_0$ (km/sec/Mpc) & - & $75.3_{-2.6}^{+1.3}$ & $71.9 \pm 2.7$ \\
      \hline
    \end{tabular}
    \caption{Mean values and 68\% C.L. errors of the cosmological
      parameters obtained from the analysis of WMAP5 alone, WMAP5 + HST,
   and WMAP5 + BBN in the lepton asymmetric universe with massless
   neutrinos. For the
   constraints on $\xi$, 95\%
  C.L. errors are also shown, and the upper bounds on $N_{\rm eff}$,
  and $Y_p$ are at 95\% C.L. Here we consider the case $\xi_e = \xi_{\mu,\tau}$ and use
   the relation $Y_p = Y_p^{\rm BBN}$. Three parameters at the bottom are
   the derived parameters from the MCMC sampling.}   
    \label{xi1m0bbnYp}
  \end{center}
\end{table}

\begin{table}[h]
  \begin{center}
    \begin{tabular}{cccc}
      \hline
      parameter & WMAP5 alone & WMAP5 + HST & WMAP5 + BBN \\
      \hline \hline
      $100 \omega_b$ & $2.257^{+0.062}_{-0.063}$ & $2.249^{+0.062}_{-0.061}$ &
      $2.253 \pm 0.059$ \\
      $\omega_{dm}$ & $0.1614^{+0.0133}_{-0.0256}$ & $0.1343^{+0.0063}_{-0.0139}$ &
      $0.1256^{+0.0059}_{-0.0097}$ \\ 
      $\xi_e$ & $1.22^{+0.58}_{-0.64}{}^{+1.74}_{-1.61}$ &
      $0.74^{+0.43}_{-0.52}{}^{+1.46}_{-1.26}$ &
      $0.040^{+0.047}_{-0.045}{}^{+0.093}_{-0.070}$ \\
      $|\xi_{\mu,\tau}|$ & $< 2.35$ & $< 1.78$ & $< 1.67$ \\
      \hline
      $N_{\rm eff}$ & $< 10.7$ & $< 7.13$ & $< 5.78$ \\
      $Y_p$ & $< 0.38$ & $< 0.41$ & $0.250 \pm 0.004$ \\
      $H_0$ (km/sec/Mpc) & - & $77.8_{-2.9}^{+2.1}$ & $76.4^{+1.9}_{-2.7}$ \\
      \hline
    \end{tabular}
\caption{The same as TABLE 4
, but for the case with $\xi_e\neq \xi_{\mu,\tau}$.} 
\label{xi2m0bbnYp}
  \end{center}
\end{table}

\bibstyle{JHEP}
\bibliography{paper}
\nocite{*}
\end{document}